\documentclass[conference]{IEEEtran}

\ifCLASSOPTIONcompsoc
  \usepackage[nocompress]{cite}
\else
  \usepackage{cite}
\fi
\usepackage[hidelinks]{hyperref}
\usepackage{orcidlink}
\usepackage{amsmath,amssymb,amsfonts}
\usepackage[ruled,vlined,linesnumbered,noend]{algorithm2e}
\usepackage{orcidlink}
\usepackage{graphicx}
\usepackage{tabularx}
\usepackage{float}
\usepackage{makecell}
\usepackage{bbding}
\usepackage{centernot}
\usepackage{tablefootnote}
\SetKw{Continue}{continue}

\usepackage{booktabs}
\usepackage{subcaption}
\usepackage{textcomp}
\usepackage{xcolor}

\begin{document}

\title{Dynamic Service Scheduling and Resource Management in Energy-Harvesting Multi-access Edge Computing}

\author{
    \IEEEauthorblockN{
        Shuyi Chen\IEEEauthorrefmark{1}\IEEEauthorrefmark{2}\orcidlink{0000-0003-0745-4083},
        Panagiotis Oikonomou\IEEEauthorrefmark{3}\orcidlink{0000-0002-5564-2591}, 
        Zhengchang Hua\IEEEauthorrefmark{1}\IEEEauthorrefmark{2}\orcidlink{0000-0002-3970-6129}, 
        Nikos Tziritas\IEEEauthorrefmark{3}\orcidlink{0000-0002-2091-2037},\\
        Karim Djemame\IEEEauthorrefmark{1}\orcidlink{0000-0001-5811-5263}, 
        Nan Zhang\IEEEauthorrefmark{2}\orcidlink{0000-0002-5728-0440} and 
        Georgios Theodoropoulos\IEEEauthorrefmark{4}\IEEEauthorrefmark{2}\orcidlink{0000-0002-7448-5886}
    }
    
    \IEEEauthorblockA{\IEEEauthorrefmark{1}School of Computer Science,University of Leeds, Leeds, UK}
    \IEEEauthorblockA{\IEEEauthorrefmark{2}Department of Computer Science and Engineering, Southern University of Science and Technology (SUSTech), Shenzhen, China}
    \IEEEauthorblockA{\IEEEauthorrefmark{3}Department of Informatics and Telecommunications, University of Thessaly, Lamia, Greece}
    \IEEEauthorblockA{\IEEEauthorrefmark{4}Research Institute of Trustworthy Autonomous Systems (RITAS), SUSTech, Shenzhen, China}
    
    \IEEEauthorblockA{\IEEEauthorrefmark{1}\{scsche,sczh,k.djemame\}@leeds.ac.uk~\IEEEauthorrefmark{3}\{paikonom, nitzirit\}@uth.gr~\IEEEauthorrefmark{2}zhangn2019@sustech.edu.cn~\IEEEauthorrefmark{4}theogeorgios@gmail.com}
}

\maketitle

\begin{abstract}
Multi-access Edge Computing (MEC) delivers low-latency services by hosting applications near end-users. To promote sustainability, these systems are increasingly integrated with renewable Energy Harvesting (EH) technologies, enabling operation where grid electricity is unavailable. However, balancing the intermittent nature of harvested energy with dynamic user demand presents a significant resource allocation challenge. This work proposes an online strategy for an MEC system powered exclusively by EH to address this trade-off. Our strategy dynamically schedules computational tasks with dependencies and governs energy consumption through real-time decisions on server frequency scaling and service module migration. Experiments using real-world datasets demonstrate our algorithm's effectiveness in efficiently utilizing harvested energy while maintaining low service latency.
\end{abstract}
\begin{IEEEkeywords}
Multi-access edge computing, energy harvesting, online scheduling, DAG
\end{IEEEkeywords}
\section{Introduction}
Multi-access Edge Computing is a transformative paradigm that embeds computing capabilities at the network edge to provide low-latency services for mobile subscribers. By connecting to a Micro Datacenter (MDC)'s associated base station through the Radio Access Network (RAN), users can offload computationally intensive tasks for rapid processing \cite{oikonomou2021use}. Common use cases include real-time video analytics, connected vehicle communications, and interactive mobile gaming. While powerful, this distributed model faces significant constraints, including limited computational resources, fluctuating energy availability, and constant user mobility. Effectively managing these limited, geographically dispersed resources to deliver a stable Quality of Service (QoS) in such a dynamic environment remains a key challenge.

Integrating Energy Harvesting is a key strategy for sustainable, low-carbon edge computing\cite{nandi2024task}. Instead of relying solely on the grid, EH devices capture ambient energy from various sources. Common examples include solar panels on outdoor edge nodes and miniature wind turbines for deployments in open areas \cite{yang2022dynamic}. By harnessing renewable sources, EH devices can power edge systems with clean energy, enabling self-sufficient operation in remote regions where traditional power infrastructure is unavailable or unreliable\cite{lu2021green}.

\begin{figure*}[htbp]
\begin{minipage}[b]{.32\textwidth}
\begin{subfigure}{\linewidth}
\centering
\includegraphics[width=0.99\textwidth]{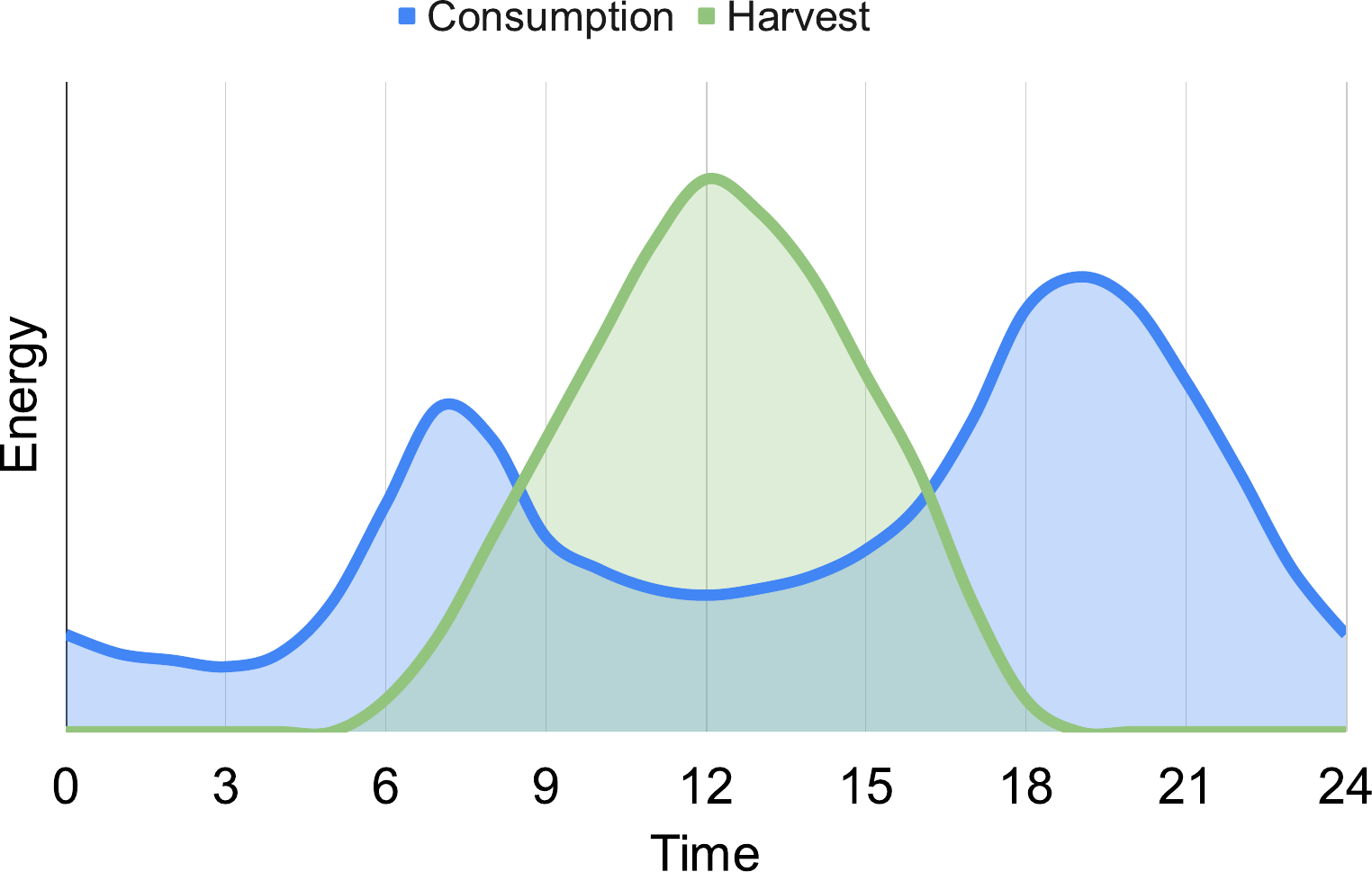}
\caption{Mismatched energy harness and consumption}
\label{fig:mismatch}
\end{subfigure}
\end{minipage}
\hfill
\begin{minipage}[b]{.33\textwidth}
\begin{subfigure}{\linewidth}
\centering
\includegraphics[width=0.99\textwidth]{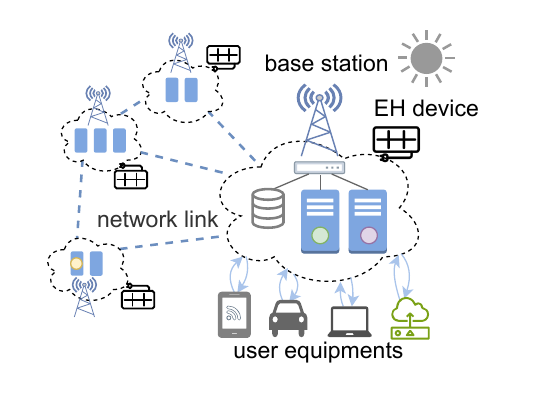}
\caption{EH-MEC network example}
\label{fig:mec-nw}
\end{subfigure}
\end{minipage}
\hfill
\begin{minipage}[b]{.33\textwidth}
\begin{subfigure}{\linewidth}
\centering
\includegraphics[width=0.99\textwidth]{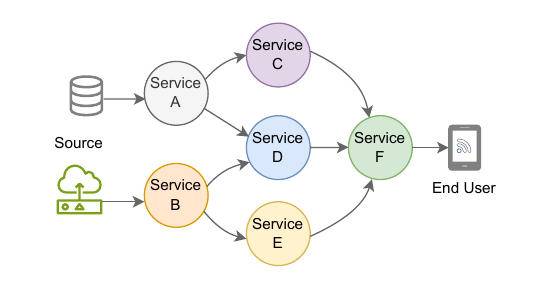}
\caption{Application graph example}
\label{fig:app}
\end{subfigure}
\end{minipage}
\caption{Examples of the EH-MEC system status, network topology and applications.}
\end{figure*}

The primary challenge lies in the stochastic nature of this energy generation, which is affected by environmental factors \cite{zhao2021deep}. This often creates a temporal mismatch between energy harvesting and consumption, as shown in Figure~\ref{fig:mismatch}. For example, solar generation during the day may not align with the usage in the evening. This discrepancy strains the battery, making it crucial to co-manage task scheduling and energy allocation for reliable system operation.

A fundamental challenge for MEC providers is the complex resource allocation and scheduling problem. Providers must intelligently assign limited resources to minimise operational costs like energy consumption \cite{premsankar2022energy} while meeting application performance requirements. This creates a difficult trade-off, as power-saving techniques like Dynamic Voltage and Frequency Scaling (DVFS) must be balanced against computational demands \cite{kazemi2025energy}.

This challenge is compounded by modern applications, which are often composed of interdependent components forming service chains \cite{liao2021dependency}. As illustrated in Figure~\ref{fig:mec-nw}, these service chains must be strategically placed across the distributed network to satisfy their dataflow requirements \cite{maray2023dependent}. Therefore, building a robust MEC system requires managing both low-level resource allocation and the high-level placement of these interconnected services.

While prior research has independently explored energy-aware scheduling for DAGs or migration in hybrid MEC systems, a unified strategy that jointly optimises both DVFS and service migration for dependency-aware applications in a purely energy-harvesting environment remains an open challenge. To address this gap, this paper introduces a novel online heuristic-based algorithm for MEC systems powered entirely by harvested energy. Our approach periodically assesses system energy dynamics to strategically adjust server frequencies and migrate service modules, aiming to balance energy efficiency and low service latency. The algorithm's efficacy was evaluated through 7,200 rounds of extensive simulations using real-world data workflows. The primary contribution of this paper is a unified online strategy that jointly manages service scheduling and energy allocation for dependency-aware applications on purely energy-harvesting MEC systems.

The remainder of this paper is organised as follows. Section~\ref{sec:related} reviews the literature and identifies the existing research gap. In Section~\ref{sec:sys-model}, we describe the system model and formulate the optimization problem. Section~\ref{sec:algo} presents our proposed algorithm in detail. We then present and discuss the experimental results in Section~\ref{sec:exp}. Finally, Section~\ref{sec:con} concludes the paper and suggests directions for future work.

\section{Related Work}\label{sec:related}
In this section, we review the literature across three key domains that inform our work: energy-efficient resource management in MEC, task scheduling for energy-harvesting systems, and dependency-aware service placement.

\textbf{Energy-Efficient Resource Management in MEC:}\
To improve energy efficiency in edge computing, resource management strategies typically balance performance and conservation, primarily through optimal service placement and dynamic adjustments. For instance, \cite{premsankar2022energy} proposed a heuristic for placing AI applications at the edge to reduce energy consumption. Similarly, \cite{yang2025energy} aims to find an optimal task offloading and bandwidth allocation policy in grid-powered MEC. \cite{li2024energy} presented an online service migration mechanism to balance the load among multiple capacity-limited IoT devices. While our work also uses migration, our strategy differs fundamentally: it is triggered by the dynamic imbalance between harvested energy and system demand, not just computational load.

Another key aspect of energy management involves Dynamic Voltage and Frequency Scaling (DVFS), a common technique to manage the trade-off between server performance and power consumption. \cite{zhao2021deep} used DVFS for resource management in multi-cloud environments and exploits the fundamental trade-off between performance and power; reducing frequency saves energy at the cost of increased latency. This trade-off is explicitly addressed by \cite{kazemi2025energy}, who proposed a task scheduling algorithm with a DVFS-based mechanism to reduce energy overhead in latency-constrained scenarios. While our strategy also uses frequency scaling, our decision-making is uniquely driven by the intermittent availability of harvested energy rather than latency constraints alone.

\textbf{Scheduling in Energy-Harvesting (EH) Systems:}\
Scheduling in energy-harvesting systems is particularly challenging due to the intermittent nature of the energy supply. Many studies have applied EH techniques to various computing domains. \cite{jayanetti2024multi} proposed a workflow scheduling framework in a multi-cloud environment powered by both brown and green energy. In the mobile domain, \cite{perin2022ease} proposed a task scheduling approach for Internet of Vehicles systems co-powered by renewables and the power grid. \cite{yang2022dynamic} designed an algorithm for optimizing energy consumption and data queue stability in UAV-enabled edge computing systems with multiple EH devices. In contrast, our work addresses the more challenging scenario of an MEC system powered exclusively by renewable energy, aiming to co-optimise user latency and task throughput under these stringent constraints.

\textbf{Dependency-Aware Workflow Scheduling:}\
Scheduling workflows with inter-dependencies adds a layer of complexity compared to scheduling monolithic or independent tasks. While \cite{yang2025energy} considers tasks with such dependencies, their focus is on single-instance execution, unlike our focus on continuous applications. In another context, \cite{zhao2021deep} manages bursts of concurrent DAG tasks rather than optimizing the performance of a single, continuous workflow. Other dependency-aware approaches, such as \cite{maray2023dependent} and \cite{liao2021dependency}, have proposed task assignment strategies in MEC without prioritizing energy constraints. These studies highlight the importance of dependency-aware scheduling but often overlook the rigorous constraints of purely renewable power, a gap our work addresses.

In summary, while prior work has separately addressed energy-aware scheduling and dependent workflows, a unified online strategy for purely renewable-powered MEC systems is lacking. Our work fills this gap with a novel algorithm that dynamically co-optimises service migration and server frequency scaling based on real-time energy availability and application structure to ensure both low latency and high service throughput.

\section{System Model and Formulation}\label{sec:sys-model}
This section presents our system model and formulates the dynamic service scheduling and resource management problem for an energy-harvesting MEC system.
\subsection{EH-MEC network model}
We consider a multi-tier EH-MEC network architecture where geo-distributed micro datacenters (MDCs) reside at the site of mobile base stations, acting as edge resource pools. As illustrated in Figure~\ref{fig:mec-nw}, the local network of each MDC contains edge servers, mobile subscribers, and source entities like IoT sensors that continuously sample data. The processing requests for the dataflows are submitted by user equipment, and the edge servers handle the computation. In each MDC, an energy harvesting (EH) device captures renewable energy from the environment and stores it in its battery. The operation of the computing servers is powered exclusively by the EH devices. This harvested power is intermittent, varying with environmental conditions like sunlight intensity.

We use $G=(M,L)$ to denote the multi-access edge network, with  $M = \left \{ m_{1},m_{2},...m_{n}\right \}$ denoting the set of MDCs, and the network links connecting the MDCs are denoted by $L = \left \{ l_{1},l_{2},...l_{p}\right \}$. The link bandwidth of each $l_{i} \in L$ is denoted by $l.bw$, with a propagation delay of $l.prop$. Each MDC $m_{i} \in M$ contains a set of edge servers $S_{i} = \left \{ s_{1},s_{2},...s_{m}\right \}$, and for each edge server $s_{i} \in S$, the maximum processing frequency of its CPU is denoted by $f_{s_{i}}^{max}$.
In the MDC local network, we use $D_{i} = \left \{ d_{1},d_{2},...d_{k}\right \}$ to represent the set of data source entities, and use $U$ to denote the user equipment. We use $EH_{m}$ to refer to the energy harvesting device powering MDC $m$, with its energy storage capacity expressed as $EH_{m}.capacity$. The power generated by each EH device is variable as it is influenced by the environment.
\subsection{Application Model}\label{app_model}
We model applications as a directed acyclic graph (DAG) $A=(V,E)$, where $V$ is the set of service modules and $E$ represents the data dependencies between them. As shown in Figure~\ref{fig:app}, a module activates only after receiving inputs from all its predecessors, and its processing results are directed to the successor modules. We assume tasks are indivisible and that a module sends results to all its successors simultaneously upon completion. For each service module $v$, the amount of workload it requires to process a single task is defined as $v.wl$, and data amount of each result packet transmitted by $v$ through a dependency edge is denoted by $v.size$.

Once an application is deployed, it can process tasks. Throughout the application's lifetime, the user triggers $A$ to process tasks continuously, but only for as long as sufficient energy is available. The processing trigger rate of $A$, $rate$, is time-varying and can follow arbitrary daily distributions to reflect dynamic user demand. We use $P\left ( v \right )$ to denote the server on which service module $v\in V$ is deployed.

\subsection{Communication and Computation Model}
In the MEC network, the communication delay for a data packet $d$ through a link $l$ is calculated as:
\begin{equation}
T_{comm}= \frac{d.size}{l.bw}+l.prop
\end{equation}
When a data packet $d$ is sent from source module $v_{src}$ to destination module $v_{dst}$, as defined in \ref{app_model}, it traverses from server $P(v_{src})$ to server $P(v_{dst})$. We use $path(v_{src}, v_{dst})$ to denote the routing path between the two servers in the network, and the delay during this process is the sum of the delays of each network hop in the path:
\begin{equation}
T_{comm}(v_{src},v_{dst}) = \sum_{l\in path(v_{src},v_{dst})}\frac{d.size}{l.bw}+l.prop
\end{equation}
For an edge server $s_{i} \in S$ that runs at frequency $f_{s_{i}}$, the execution time it needs to process a task from service $v$ can be calculated by: 
\begin{equation}\label{wl.f}
    T_{exec}(v,s_{i}) = \frac{v.wl}{f_{s_{i}}}
\end{equation}
When $n$ services are deployed on the same server, we introduce a co-location overhead: $K(n)$. Accounting for this multi-tenancy overhead, the task execution time becomes:
\begin{equation}
\label{eqn:exe2}
    T_{exec}(v|n) = T_{exec}(v) * K(n)
\end{equation}

Based on our dependency model, the Earliest Start Time (EST) for any application module v is determined by its most time-consuming predecessor. For a module $v$ with predecessors $v.pred$, its $EST$ is calculated as follows:
\begin{equation}
\label{eqn:e2e}
    EST(v) = \max_{v_{i}\in v.pred}\left \{ EST(v_{i})+T_{exec}(v_{i})+T_{comm}(v_{i},v)\right \}
\end{equation}
The end-to-end latency experienced by the user is then expressed as $LT = EST(U)$.
\subsection{Energy and Battery Model}
The energy consumption of an edge server has both static and dynamic components. We focus on the dominant dynamic portion, which is determined by the CPU's operating frequency (f) and the supply voltage (v) of its CMOS circuits. Assuming that $v$ and $f$ are linearly dependent, we derive the dynamic energy consumption of each server $s$ as:
\begin{equation}\label{eq:ec}
E_{consumed}(s) = \int \beta_{s} {f_{s}}^{3} \,dt \
\end{equation}
where $\beta_{s}$ represents a device-related energy factor.
For the purposes of monitoring and rescheduling, we model time as a sequence of discrete time slots:  $\left\{ 1, 2, 3, ..., n \right\}$, with equal lengths $\Delta t$. We use $B_{m}(k)$ to denote the battery level of the energy harvesting device $EH_{m}$ at the beginning of the $k$th time slot.
The battery's energy level in MDC $m$ is determined by the harvested energy and server consumption:
\begin{equation}\label{bk}
    B_{m}(k+1) = B_{m}(k) + E_{harvested}(m)^{k}-E_{consumed}(m)^{k}
\end{equation}
The battery level of every EH device must remain within its operational bounds:

\begin{equation}
     0 \leq B_{m}(k)\leq EH_{m}.capacity
\end{equation}
\subsection{Problem formulation}
Our objective is to minimise user latency and maximise energy efficiency, defined as the task throughput per unit of energy consumed. The solution involves finding an initial mapping of service modules to servers and then periodically adjusting this mapping using server frequency scaling and module migration. We formulate the problem as follows:\\
(i) Given an application graph $A=(V,E)$ submitted by user $U$, and the set of available servers in the EH-MEC network $G=(M,L)$, each service module $v \in V$ must be assigned to a server $s$ in $G$. Let S denote the set of all edge servers in $G$. A valid placement of all services is a mapping expressed as: $Placement:O\rightarrow S$.\\
(ii) Provided that (i) is satisfied, our optimisation goal can be formulated as follows:
\begin{equation}
\begin{split}
    P: &min\;LT, max\;  \frac{\sum A_{Completed}}{\sum_{m\in M}E_{consumed}(m)))}\\
s.t. &(i)\\
&0 \leq B_{m}(k)\leq EH_{m}.capacity, m \in M\\
&0 \leq f_{s}(k)\leq f_{s}, s \in S\\
\end{split}
\end{equation}

\section{Dynamic Scaling and Migration Algorithm}\label{sec:algo}
\begin{figure}
\centering
\includegraphics[width=0.45\textwidth]{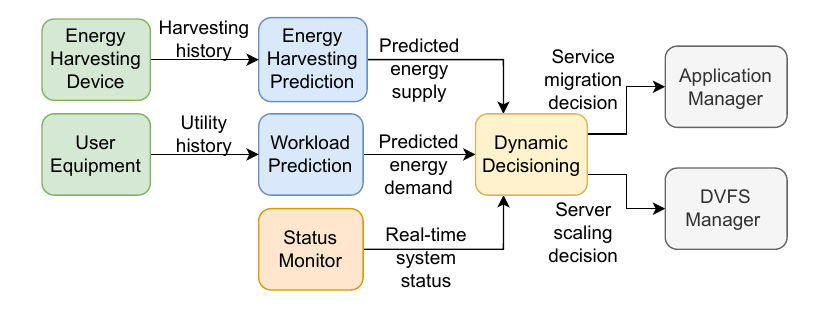}
\caption{The workflow of the dynamic algorithm proposed.}
\label{fig:scheduler}
\end{figure}

The proposed algorithm consists of two primary components: 1) an initial service assignment and 2) periodic runtime adjustments, which include server frequency scaling and service migration. The workflow of the algorithm is illustrated in Figure~\ref{fig:scheduler}.
\subsection{Initial Assignment}
The pseudo-code of this initial phase is proposed in Algorithm~\ref{alg:DSM}. This phase is guided by the Critical Path Method (CPM) and proceeds as follows: Given an application graph $A$ with its computational demands and dependencies, we first identify the critical path of application, denoted as $CP$, and then classify service modules as either \textit{critical} or \textit{non-critical}. The services composing $CP$ are assumed to have the largest impact on the end-to-end latency $LT$. Critical services, represented by $V_{Crit}$, are prioritised and assigned to Mobile Data Center (MDC) servers predicted to yield the lowest latency and energy consumption. Subsequently, non-critical services, denoted by $V_{nonCrit}$, are assigned to MDC servers under the constraint that their placement does not delay the scheduled execution of any successor nodes.
\subsection{Dynamic Resource Scheduling}
During the active runtime phase, our dynamic resource scheduler operates periodically. At the beginning of each time interval, state-of-the-art prediction methods\cite{cammarano2012pro} use historical energy harvesting and task arrival records to forecast the energy supply, denoted as                                                              $E_{supply}$, and the volume of new task arrivals for the upcoming period. Based on the current system state, the future energy demand of each Mobile Data Center (MDC), denoted as $E_{demand}[m]$, is then estimated using Equation.~\ref{eq:ec}. To facilitate resource management, MDCs are then categorised based on their predicted energy status relative to a safety threshold $\alpha$, into either a surplus group ($MDC_{sur}$) or a deficit group ($MDC_{def}$) as follows:
\begin{equation}
\begin{split}
E_{supply}[m] > E_{demand}[m]*\alpha,&\forall m \in MDC_{sur} \\
E_{supply}[m] < E_{demand}[m],&\forall m \in MDC_{def}\\
\end{split}
\end{equation}
Next, we evaluate the system's overall energy state, categorising it into one of three scenarios, each triggering a distinct policy, as shown in Algorithm~\ref{alg:DSM}:
\begin{itemize}
    \item \textit{Sufficient Scenario:} The entire system is energy-sufficient, meaning all MDCs belong to the $MDC_{sur}$ group.
    \item \textit{Deficient Scenario:} The entire system is energy-deficient, meaning all MDCs belong to the $MDC_{def}$ group.
    \item \textit{Mixed Scenario:} The system contains a combination of energy-surplus and energy-deficient nodes (i.e., both the $MDC_{sur}$ and $MDC_{def}$ groups are non-empty).
\end{itemize}

In the Sufficient Scenario, our policy enhances performance via a two-step process, detailed in Algorithm~\ref{alg:suf}:\\
\textit{Server Frequency Scaling-Up:} First, the algorithm identifies any active servers operating at sub-maximal CPU frequencies. The server CPU frequencies are then scaled up via DVFS, constrained by the energy budget and the processor's maximum physical frequency. For each server $s$ in a surplus MDC $m$, its frequency is multiplied by a scaling factor calculated as:
\begin{equation}
    scaleUpFactor = min\left( \frac{f_{s}^{max}}{f_{s}^{prev}},\sqrt{\frac{E_{supply}[m]}{\alpha*E_{demand}[m]}} \right)
\end{equation}
\textit{Critical Service Migration:} Next, the policy evaluates potential latency improvements from migrating critical services to less-loaded servers on other MDCs. A potential migration is considered a valid candidate if it both offers a latency reduction and respects the energy budget of the destination. 
A server $s^{*}$ is considered a valid migration target for a service $v$ if it satisfies two conditions. First, the predicted energy state of the target MDC, m, must be in surplus:
\begin{equation}
E_{supply}[m]^{*} > E_{demand}[m]^{*}
\end{equation}
Second, the migration must result in a lower predicted end-to-end latency:
\begin{equation}
EST(U)^{*} < EST(U)
\end{equation}
Among all valid options, the migration plan offering the greatest latency reduction is selected and denoted as $BestMigTarget$, and the decision is sent to the platform's application manager for execution.

In the Deficient Scenario, where every MDC in the system faces an energy shortfall, the algorithm adopts a conservative, energy-preservation policy, as described in Algorithm~\ref{alg:def}. The primary objective is to maintain system operation without depleting the limited energy supply. All service migrations are suspended in this scenario. Concurrently, the CPU frequency of all active servers is scaled down to a level that ensures the projected energy consumption remains within the predicted budget. These decisions are then sent to the system's DVFS manager for execution. For each server $s$ in a deficient MDC $m$, its CPU frequency will be multiplied by a factor calculated as:
\begin{equation}
    scaleDownFactor =  \sqrt{\frac{E_{supply}[m]}{E_{demand}[m]}} 
\end{equation}

The Mixed Scenario is characterised by an energy imbalance across the system, which contains MDCs from both the $MDC_{sur}$ and $MDC_{def}$ groups. The policy aims to resolve this imbalance by reallocating resources from surplus to deficit nodes through a two-phase process detailed in Algorithm~\ref{alg:mix}.\\
\textit{Phase 1.} Alleviating Deficits by Migrating Non-Critical Services:
To relieve energy-deficient MDCs ($MDC_{def}$), this phase offloads their non-critical workloads, which have a lower impact on end-to-end latency. First, all non-critical services on these MDCs are identified and sorted by computational demand for sequential migration consideration. The algorithm then sequentially attempts to migrate each service to a suitable server within the surplus group ($MDC_{sur}$ that has sufficient capacity. After each successful migration, the energy states of the source and destination MDCs are re-evaluated, potentially changing their group status. This process continues until all non-critical services from the original deficit list have been considered.\\
\textit{Phase 2.} Handling Critical Services on Remaining Deficit MDCs:
After the non-critical migrations, the policy addresses any critical services still located on MDCs that remain energy-deficient. For each of these critical services, two competing actions are evaluated to determine the optimal outcome:
\begin{itemize}
    \item Option A (Migration): Find the best possible migration target server within the $MDC_{sur}$ group.
    \item Option B (Frequency Scaling): Calculate the reduced CPU frequency at which the service's current host server must operate to stay within its local energy budget.
\end{itemize}
The algorithm then estimates the resulting end-to-end latency $EST(U)^{*}$ of the critical path for both options. The action that leads to the lower latency is selected and executed (either as a migration command or a DVFS request). The system's state is updated before proceeding to the next critical service on a deficit node.

\begin{algorithm}
\SetAlgoNoLine
\LinesNumbered
\caption{Dynamic scaling \& migration algorithm (DSM)\label{alg:DSM}}
\KwData{Application $A = (V,E)$, EH-MEC network $G=(M,L)$}
\KwResult{Server placement map $P:V\rightarrow S$, scaling and migration decisions}
\textit{\textbf{1: Initial service assignment:}}\\
Initialise $P:V\rightarrow S$, critical path $CP$;\\

\For{$v \in CP$}{
 Find MDC $m^*$ that minimises $EST(v)$;\\
Select s in $m^*.servers$ with min $EC(v,s)$;\\
    $P(v) \leftarrow s$;
}
$V_{nonCrit} \leftarrow V \setminus CP$;\\

\For{$v \in V_{nonCrit}$}{
 Find MDC $m^*$ that $EST(v)+T_{comm}(v,v.succ) \le EST(v.succ)$;\\
Select $s$ in $m^*.servers$ with min $EC(v,s)$;\\
    $P(v) \leftarrow s$;
}
\textit{\textbf{2: Server scaling and service migration:}}\\
\While{$A.active$}{

Estimate $E_{supply}, E_{demand}$ from predictions;\\

Determine $MDC_{sur}$, $MDC_{def}$;\\

$M.scenario$ = Assess($M$);\\
\uIf{M.scenario$ == Sufficient$}{
   Apply the \textit{Sufficient Policy};
  }
  \uElseIf{M.scenario $== Deficient$}{
   Apply the \textit{Deficient Policy};
}
\uElseIf{M.scenario$ == Mixed$}{
    Apply the \textit{Mixed Policy};
  }
}
\end{algorithm}

\begin{algorithm}
\SetAlgoNoLine
\LinesNumbered
\caption{Sufficient Policy\label{alg:suf}}
 \For{$m \in MDC_{sur}$}{
    $E_{sur} \leftarrow E_{supply}[m] - E_{demand}[m]$;\\
    \For{$s \in m.servers$}{
    \uIf{$s.freq < s.max\_freq$}{
    Calculate $scaleUpFactor$ with $m, E_{sur}$;\\
    $f_{s_{new}} \leftarrow f_s \times scaleUpFactor$;\\

    }
    }
    }
    \For{$v \in CP$}{
        $targetServer$ $\leftarrow$ BestMigTarget($v, MDC_{sur}$);\\
        \uIf{targetServer}{
        $P(v) \leftarrow$ TargetServer;\\
		Migrate($v,TargetServer$);\\
        }
    }
\end{algorithm}

\begin{algorithm}
\SetAlgoNoLine
\LinesNumbered
\caption{Deficit Policy\label{alg:def}}
 \For{$m \in MDC_{def}$}{
    $E_{def} \leftarrow E_{demand}[m] - E_{supply}[m]$;\\
    \For{$s \in m.servers$}{
    Calculate $scaleDownFactor$ with $m, E_{def}$;\\
    $f_{s_{new}} \leftarrow f_s \times scaleDownFactor$;\\
 
    }
    }
  
\end{algorithm}

\begin{algorithm}
\SetAlgoNoLine
\LinesNumbered
\caption{Mixed Policy\label{alg:mix}}
\For{$m \in MDC_{def}$}{
     $V_{migrate} \leftarrow \left\{ v | P(v) = m, v \notin CP\right\}$;\\
     Sort $V_{migrate}$ by consumption;\\
     \For{$v \in V_{migrate}$}{
     $tgtServer \leftarrow$ BestMigTarget($v, MDC_{sur}$);\\
     $P(v) \leftarrow targetServer$;\\
		Migrate($v,TargetServer$);\\}}
     \For{$m \in MDC_{def}$}{
        $V_{critical\_deficit} \leftarrow \left\{v | P(v) = m, v \in CP\right\}$;\\
        \For{$v \in V_{critical\_deficit}$}{
        $tgtServer \leftarrow$ BestMigTarget($v, MDC_{sur}$);\\
    Calculate $scaleDownFactor$ with $m, E_{def}$;\\
        Calculate $EST_{mig}$ and $EST_{sca}$;\\
        \eIf{$EST_{mig}>EST_{sca}$}{
    $f_{s_{new}} \leftarrow f_s \times scaleDownFactor$;\\
        }{Migrate($v,targetServer$);}
        }
     }
\end{algorithm}

\section{Experimental Evaluation}\label{sec:exp}
\subsection{Performance Indicators \& Setup}
\begin{figure}
\centering
\includegraphics[width=0.45\textwidth]{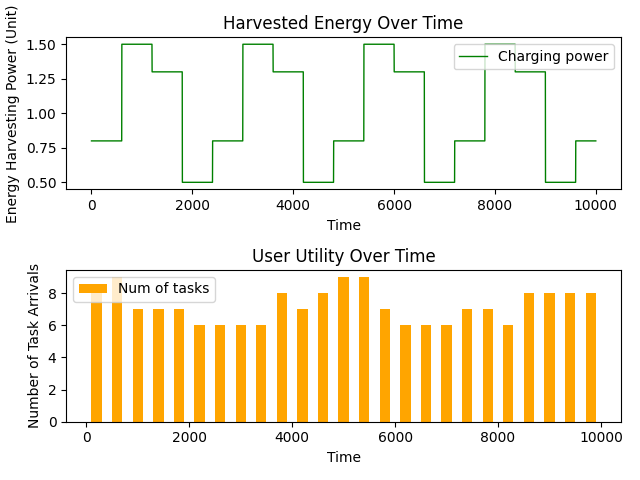}
\caption{The energy harvesting power and amount of incoming tasks over time.}
\label{fig:utility}
\end{figure}

To evaluate the proposed scheduling scheme, we used and extended the YAFS fog simulator \cite{lera2019yafs} to support the sequential processing of dependent tasks and the dynamic harvesting of energy. Network topologies were generated using the Barabasi-Albert model. Each MDC is powered exclusively by an energy harvesting device with battery storage, featuring time-varying harvesting power as shown in Figure~\ref{fig:utility}. To prolong the operational lifespan of the battery, a safe line energy threshold is defined at 10\%; i.e., when the battery level drops below this threshold, the corresponding MDC becomes non-operational until sufficient energy is harvested and the battery is recharged. Regarding MEC characteristics, the number of MDCs, denoted by $n$, was varied as $n \in \left\{ 2,4,8 \right\}$. Similarly, the number of servers, denoted by m, within each MDC was varied as $m \in \left\{ 1,2,4\right\}$. The computing resources are heterogeneous, with their CPU frequencies uniformly distributed between 1.5 GHz and 2.5 GHz. The propagation delay is set to 5 ms, the bandwidth to 1 Gbps, and the device power factor to $10^{-27}$.

To simulate different environmental conditions, the charging multiplier, denoted by $cm$ (i.e., the rate of energy harvesting), was varied among the values {0.75,1,1.25,1.5}. Lower values indicate limited sunlight or weak wind, while larger values correspond to stronger sunlight or wind conditions. The maximum capacity for each battery was set to 3600 J, and the initial battery level for each MDC was randomly set within the range of 25\% to 75\% of this maximum. To make our results more general and realistic, we used real-world workloads from the Alibaba cluster trace dataset.
This dataset includes information about DAGs from actual workloads running on a large data center, including services like video streaming and machine learning inference \cite{oikonomou2018scheduling}. For our experiments, we randomly chose 10 applications from the dataset that had more than 10 modules and present their main characteristics in Table~\ref{tab:apps}.
\begin{table}
\centering
\begin{tabular}{@{}cccccc@{}}
\toprule
Job id & \multicolumn{1}{c}{\textit{$|V|$}} & \multicolumn{1}{c}{$|E|$} & \multicolumn{1}{c}{\makecell{Max \\ degree}} & \multicolumn{1}{c}{\makecell{Average \\ transfer volume}} & \multicolumn{1}{c}{\makecell{Average \\ workload}} \\ \midrule
1 & 12 & 9 & 3 & 39.33 & 9.50 \\
2 & 16 & 17 & 6 & 29.00 & 257.88 \\
3 & 16 & 17 & 7 & 23.63 & 40.50 \\
4 & 17 & 17 & 5 & 42.82 & 13.53 \\
5 & 16 & 17 & 3 & 46.06 & 35.75 \\
6 & 12 & 11 & 6 & 38.67 & 8.17 \\
7 & 10 & 10 & 4 & 44.30 & 14.40 \\
8 & 10 & 9 & 5 & 35.60 & 7.10 \\
9 & 16 & 16 & 2 & 47.19 & 1.00 \\
10 & 10 & 9 & 4 & 43.40 & 32.70 \\ 
\bottomrule
\end{tabular}

\captionsetup{justification=centering}
\captionof{table}{Application characteristics}
\label{tab:apps}
\end{table}

To evaluate our proposed strategy, we manage to identify suitable external baselines. Our work addresses a specific, multi-faceted problem that combines 1) DAG-based task dependencies, 2) exclusive reliance on harvested energy, and 3) the joint, online optimization of both DVFS and service migration. While existing research addresses subsets of these challenges, we found no contemporary schedulers that holistically integrate all of these components. Adapting algorithms from these related but distinct domains would require significant modifications, potentially leading to an unfair comparison.

\begin{figure*}[htb]
\centering
  \begin{minipage}[t]{.4\textwidth}
\begin{subfigure}[t]{\linewidth}
\centering
        \includegraphics[width=0.99\linewidth]{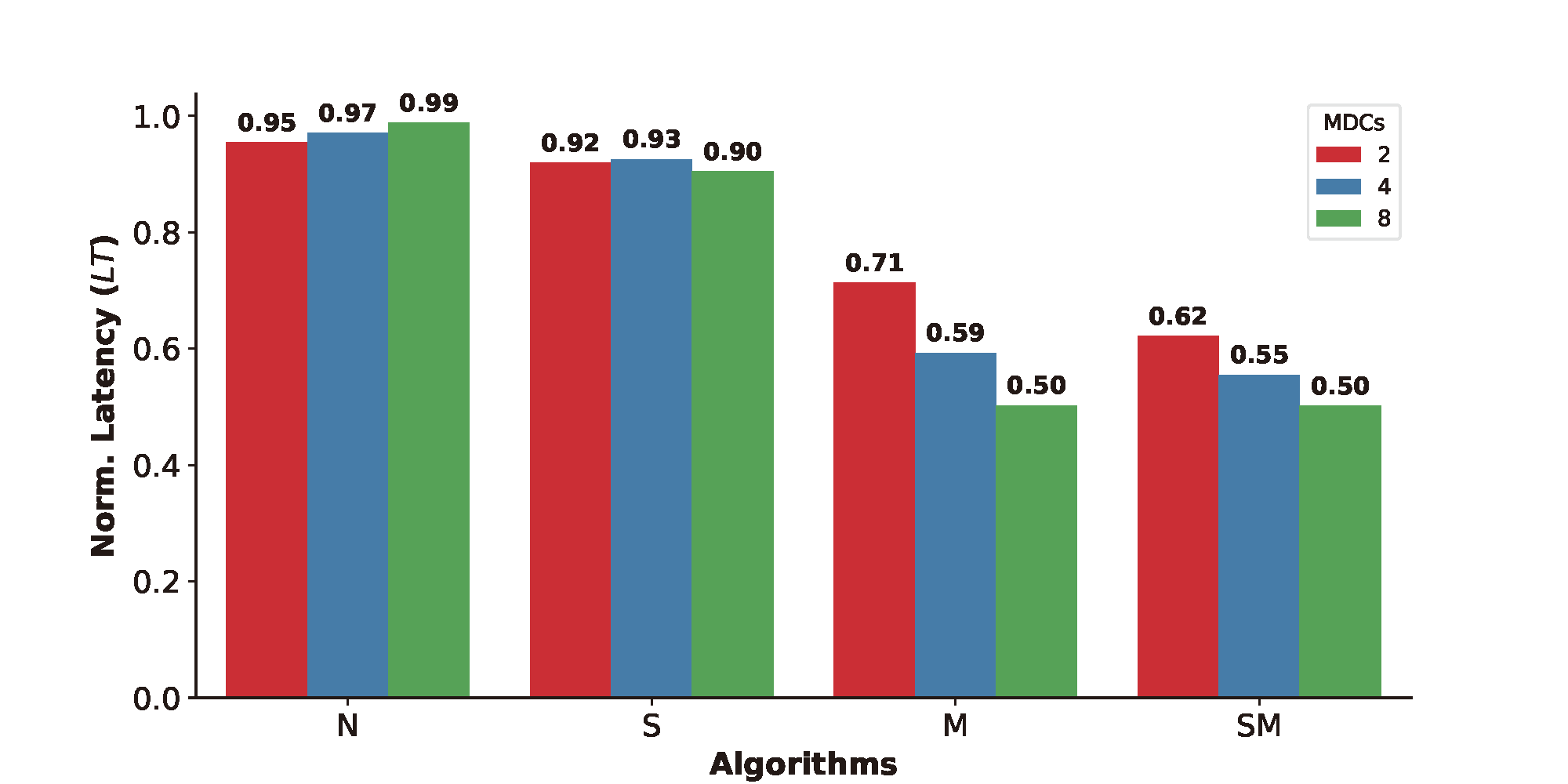}
        \caption{$LT$, $m \in [2,4,8]$}
        \label{fig:norm_latency_plot_MDCs}
    \end{subfigure}
   \vfill
      \begin{subfigure}[t]{\linewidth}
      \centering
    \includegraphics[width=0.99\linewidth]{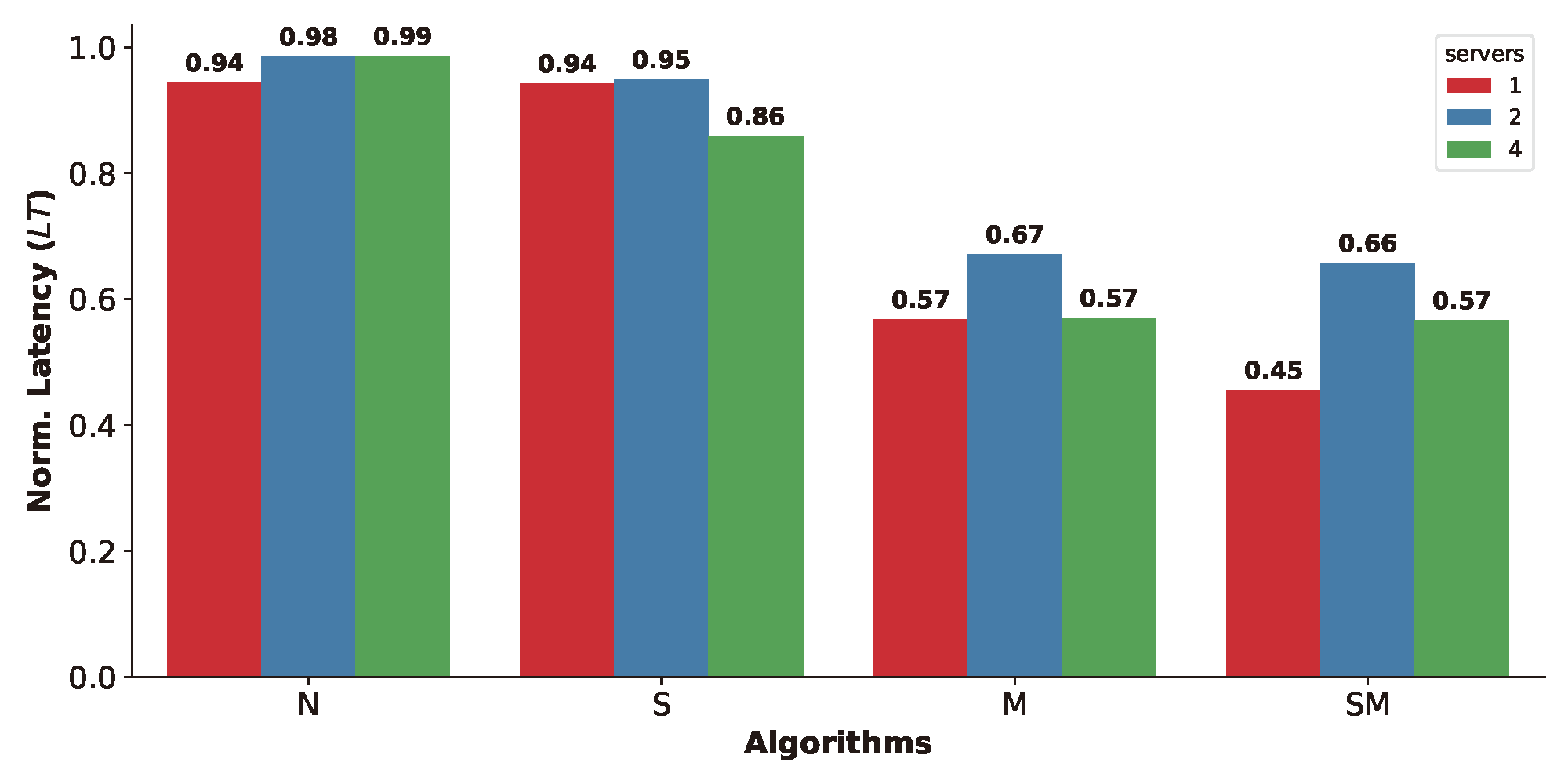}
        \caption{$LT$, $s \in [1,2,4]$}
    \label{fig:norm_latency_plot_servers}
\end{subfigure}
      \end{minipage}
  \begin{minipage}[t]{.4\textwidth}
  \begin{subfigure}[t]{\linewidth}
  \centering
        \includegraphics[width=0.99\linewidth]{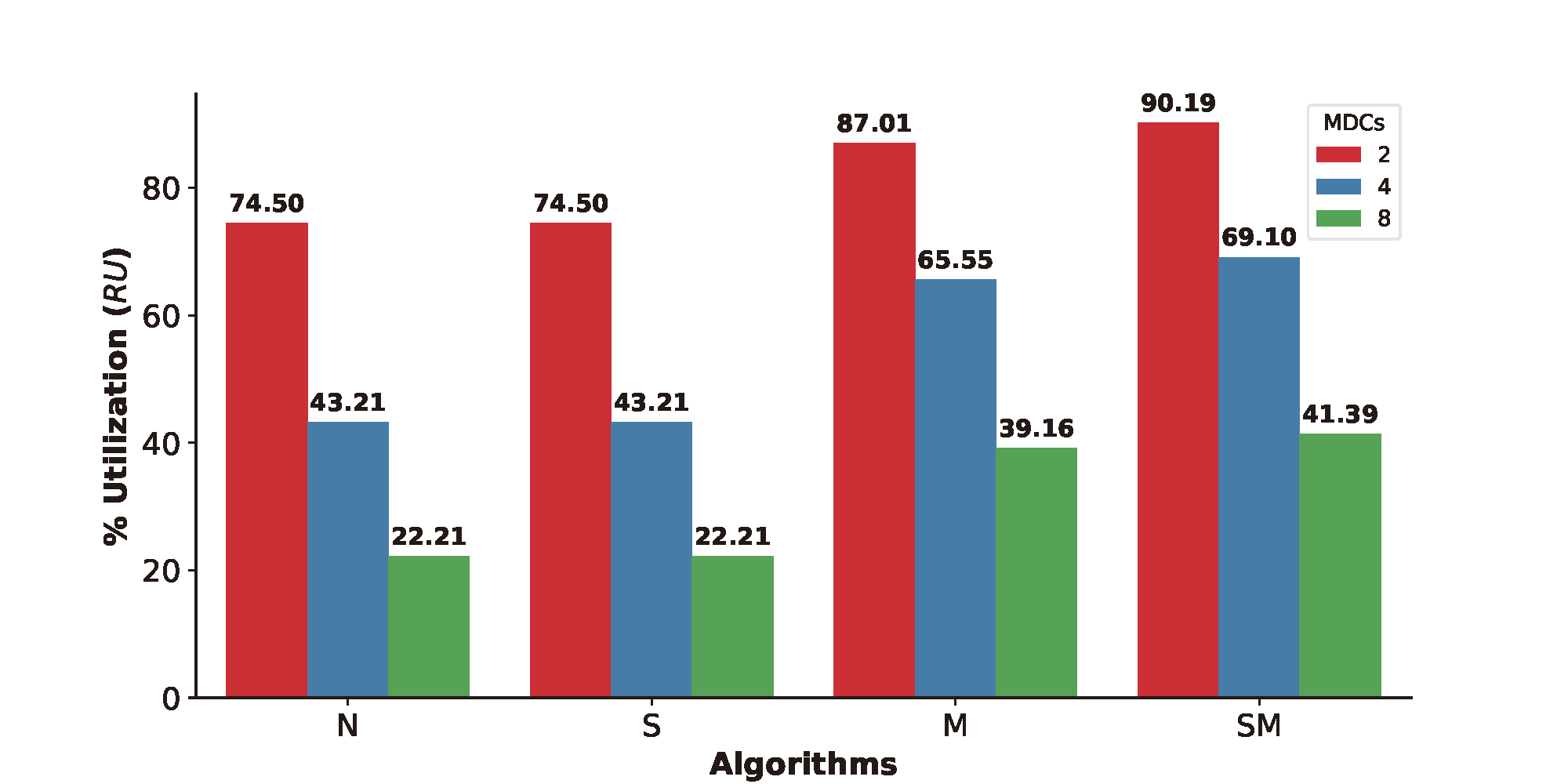}
        \caption{$RU$, $m \in [2,4,8]$}
        \label{fig:util_plot_MDCs}
    \end{subfigure}
    \vfill
      \begin{subfigure}[t]{\linewidth}
      \centering
    \includegraphics[width=0.99\linewidth]{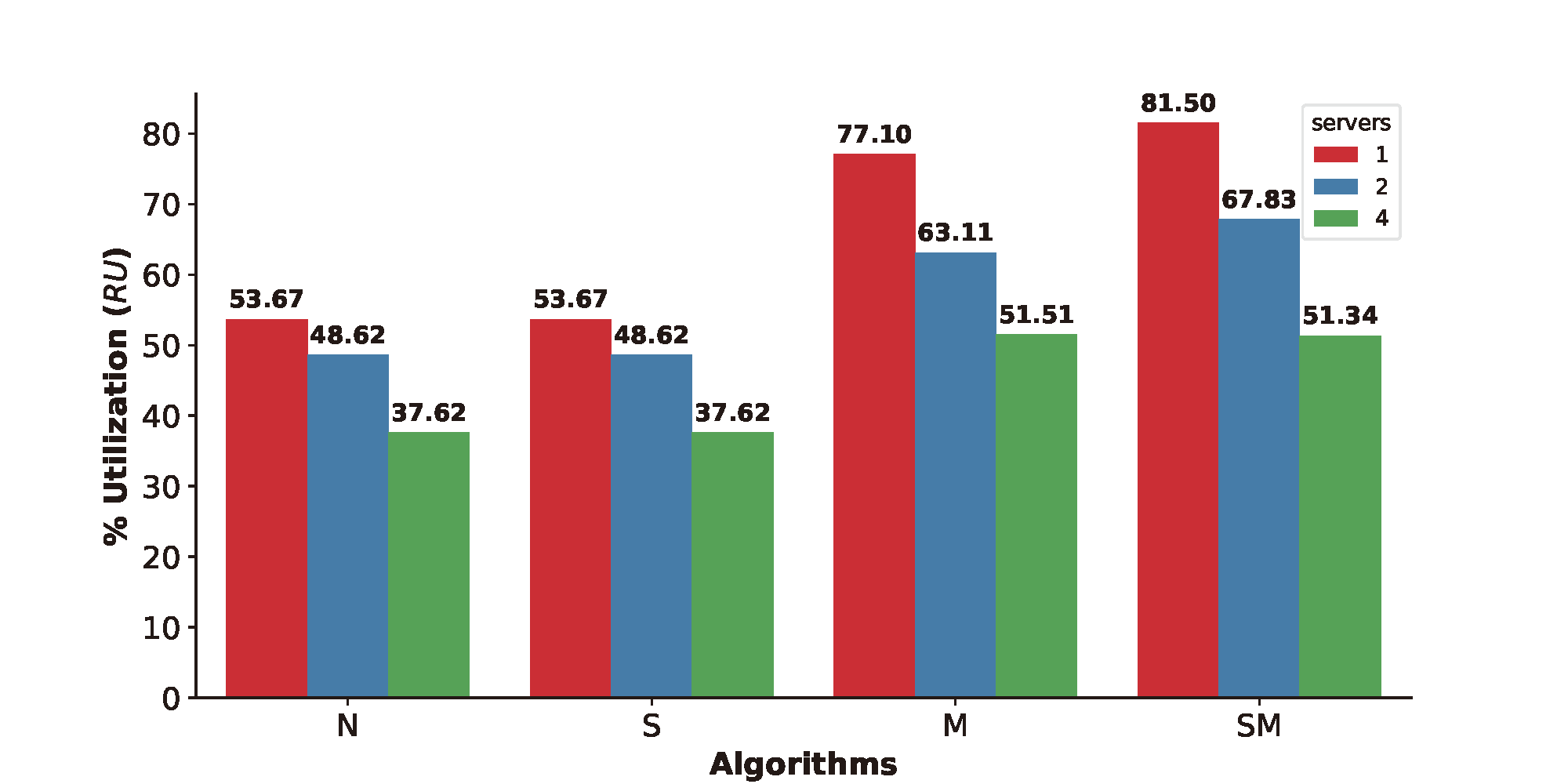}
        \caption{$RU$, $s \in [1,2,4]$}
    \label{fig:util_plot_servers}
\end{subfigure}
  \end{minipage}

    \caption{Results on latency and utilisation}
\end{figure*}

To clearly isolate and quantify the specific contributions of each component of our strategy, we adopt an ablation study approach, and compare four different scheduling policies. The first policy, denoted as $N$, takes no adaptive action after the initial placement of the services. The second policy, $S$, only applies dynamic scaling of server frequencies. The third policy, $M$, selectively migrates specific services to servers within the same or different MDCs. Finally, the $SM$ policy combines both scaling and migration strategies. The initial service placement on MDCs is performed using the EDEM algorithm  \cite{chen2024efficient,chen2025qos}. This approach allows for a clear and unambiguous analysis of how each adaptive strategy contributes to system performance under the stringent constraints of a purely EH-MEC environment.

Each experimental configuration was simulated over 10,000 global time steps and repeated 5 times to ensure reliable average results. Tasks arrive continuously for processing within the MDCs, following the pattern illustrated in Figure~\ref{fig:utility}. Scheduling decisions are made every 100 global time steps to adapt to the continuous arrival of new data from IoT sensors. The simulations were conducted on a server with a 4x Intel Xeon Gold 6230N CPU, 256 GB of RAM, and the Ubuntu 20.04 operating system.

\begin{figure*}[htbp]
    \centering
    \begin{subfigure}{.45\linewidth}
    \centering
        \includegraphics[width=0.99\linewidth]{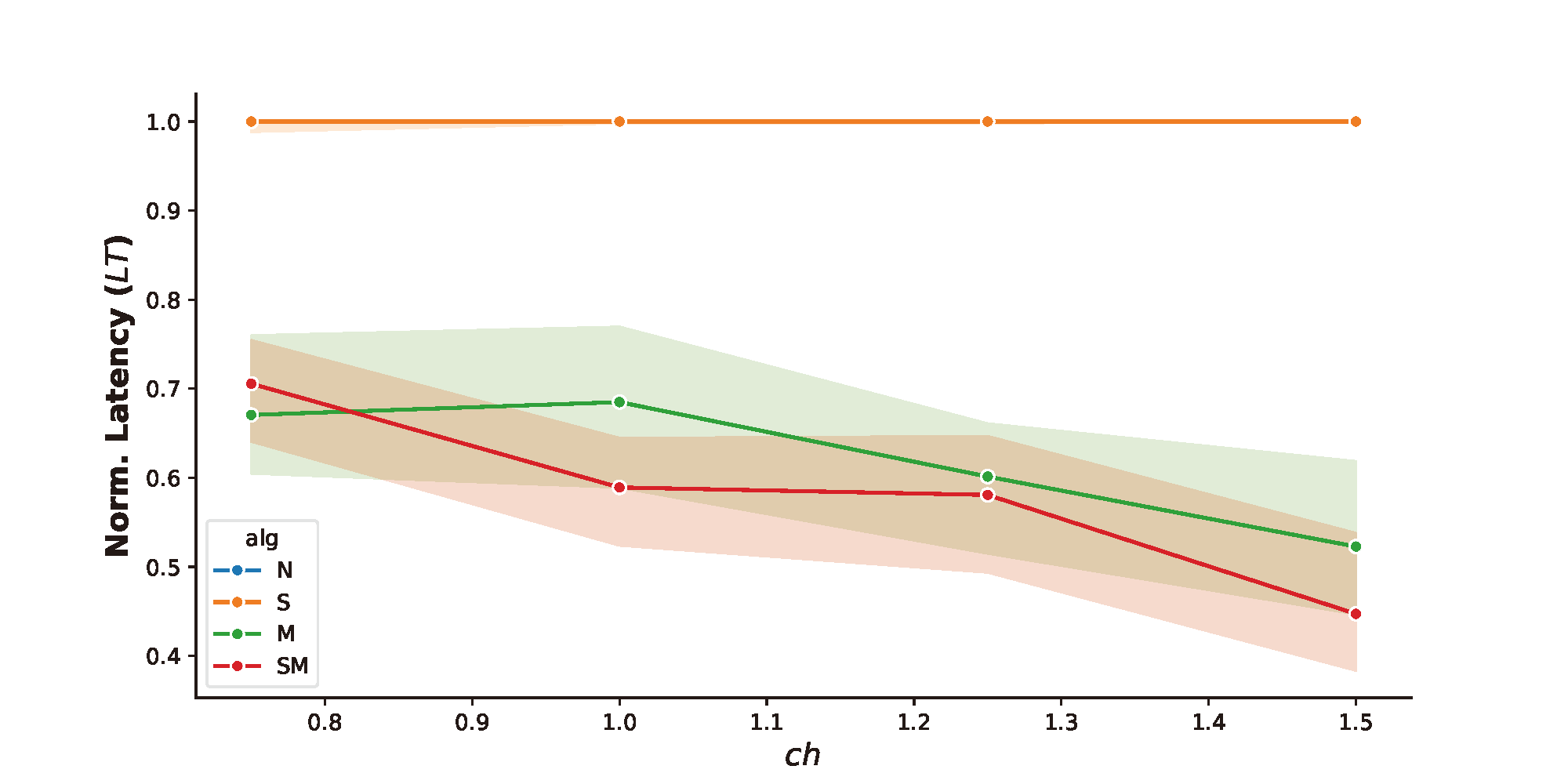}
        \caption{$LT$, $ch \in [0.75,1,1.25,1.5]$}
        \label{fig:norm_latency_plot_chMult}
    \end{subfigure}
  \begin{subfigure}{.45\linewidth}
  \centering
    \includegraphics[width=0.99\linewidth]{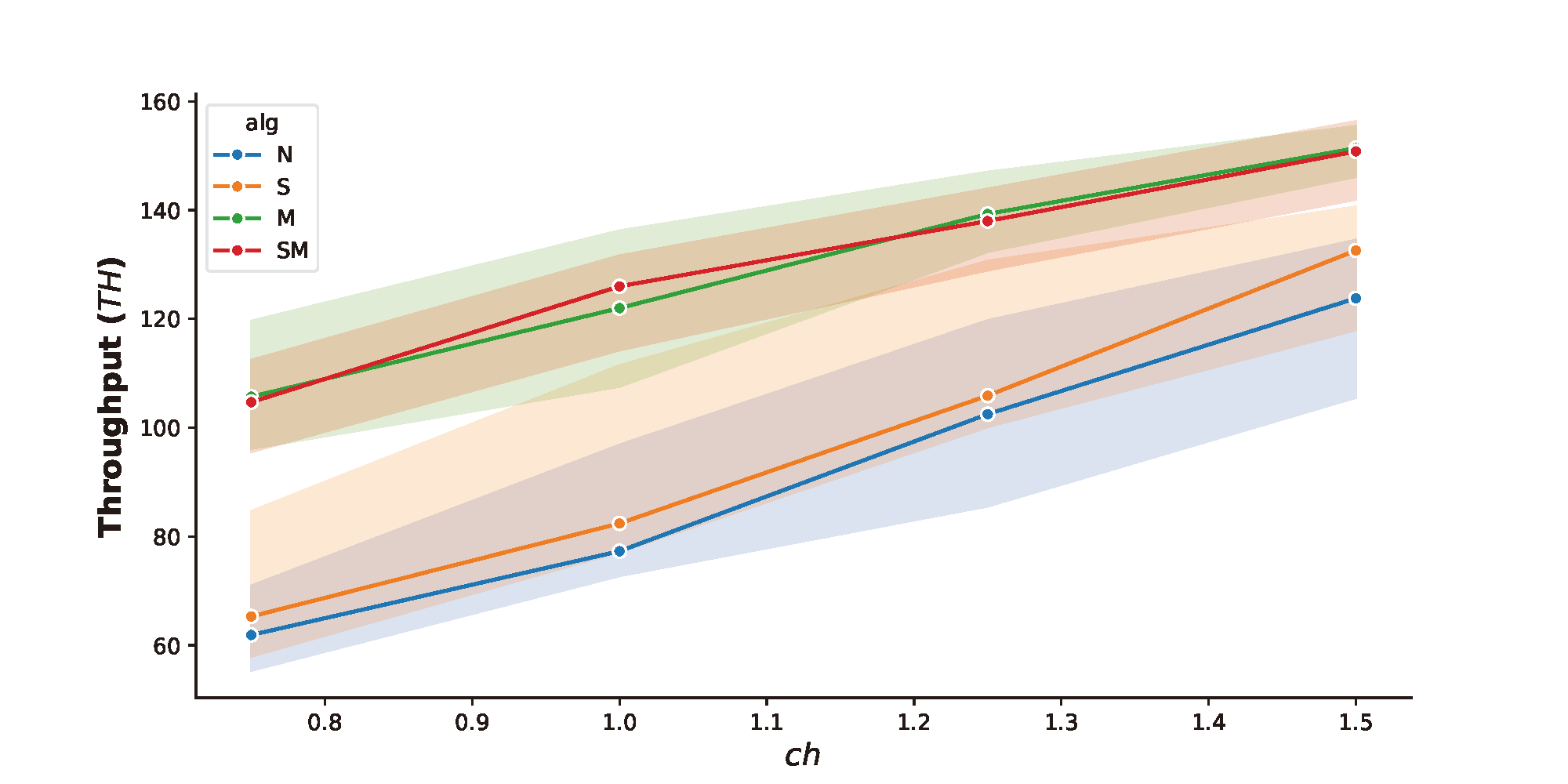}
        \caption{$TH$, $ch \in [0.75,1,1.25,1.5]$}
    \label{fig:throughput_plot_chMult}
\end{subfigure}

    \begin{subfigure}{.4\linewidth}
        \centering
        \includegraphics[width=0.99\linewidth]{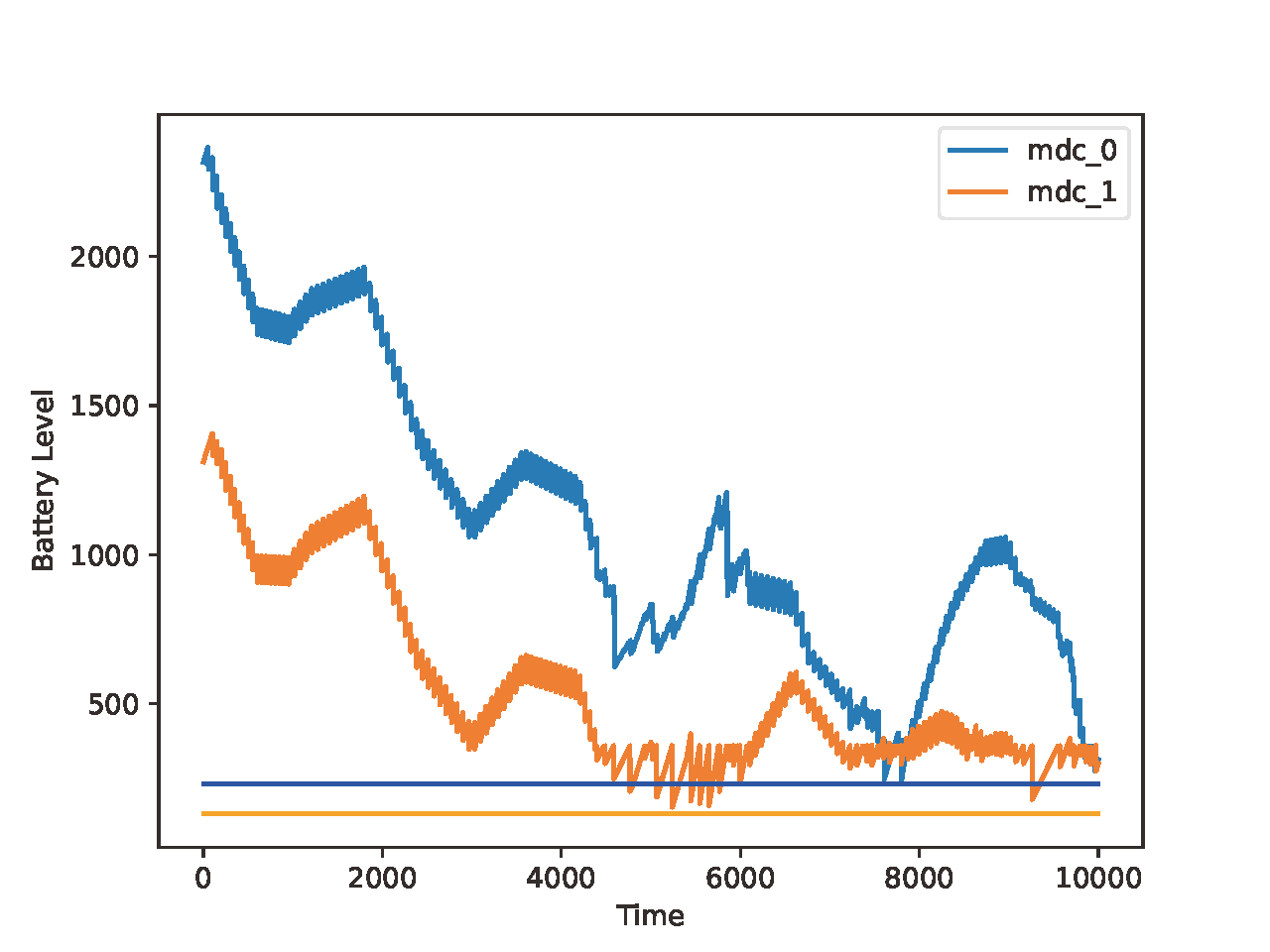}
        \caption{Battery levels per timestep}
        \label{fig:mdc_battery_levels_example}
    \end{subfigure}
    \begin{subfigure}{.4\linewidth}
        \centering
        \includegraphics[width=0.99\textwidth]{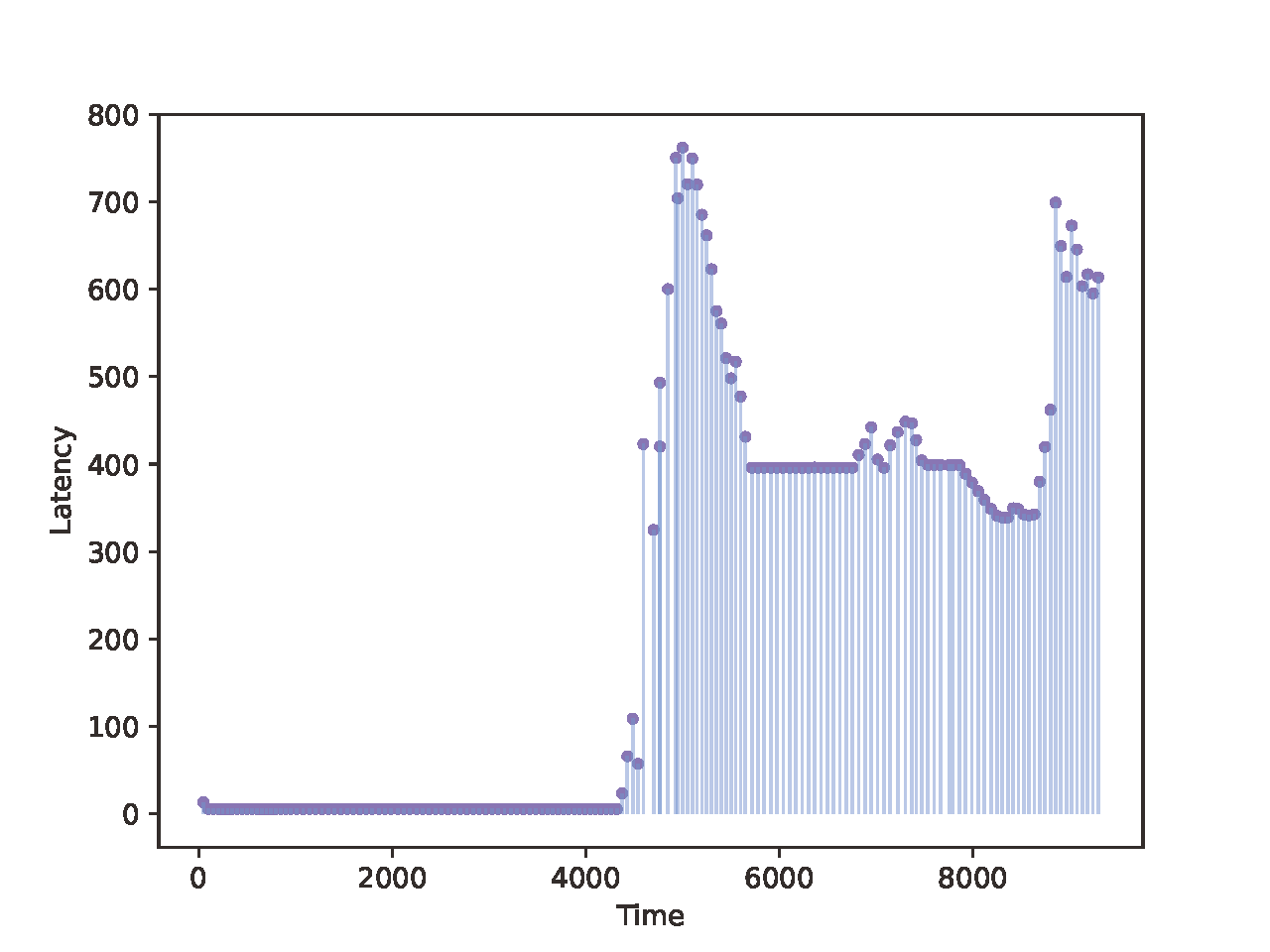}
        \caption{Latency values per timestep}
        \label{fig:latency_distribution_example}
    \end{subfigure}
    
\caption{Figure caption}
\label{}
\end{figure*}

We evaluate the performance of the algorithms using the following three metrics: a) Average User-Experienced Latency ($LT$): the mean response time for processed user requests, extracted from message timestamps stored in the simulation traces, b) Throughput ($TH$): the total number of requests successfully executed, and c) Resource Utilization ($RU$), a measure of how effectively the available resources are used. A total of 7200 experiments were conducted.
\subsection{Performance Assessment}

\begin{figure}[htbp]
\centering
\includegraphics[width=0.5\textwidth]{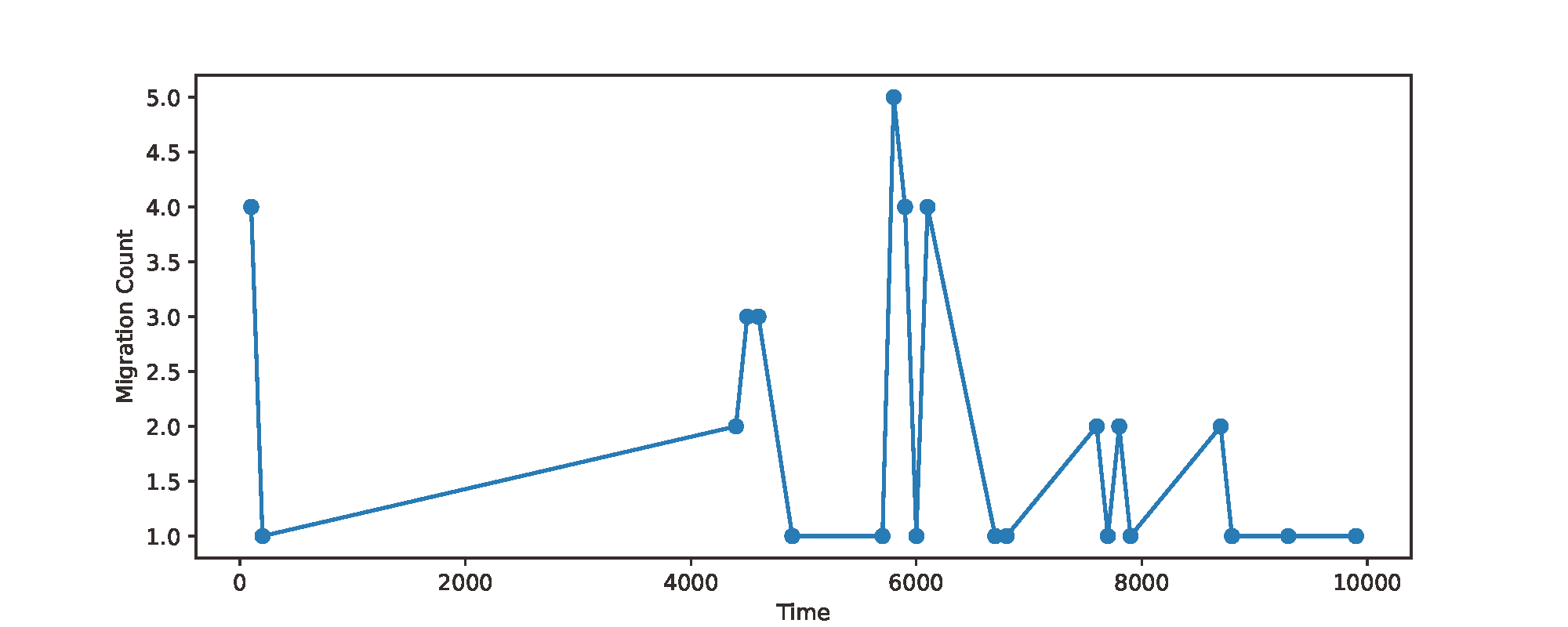}
\caption{Results on battery levels and migration}
\label{fig:Migrations_example}
\end{figure}

The results were first averaged across runs, and then $LT$ was normalised per experimental configuration (i.e., for each unique combination of application, number of MDCs, servers, and charging multiplier). This normalization was used since directly comparing raw results can be misleading. For $LT$, each latency value was divided by the maximum latency observed within its respective configuration group. This ensures that the worst-performing algorithm receives a score of 1, and better-performing ones receive proportionally smaller values.

In Figure~\ref{fig:norm_latency_plot_MDCs} and Figure~~\ref{fig:norm_latency_plot_servers}, we plot the normalised latency for all algorithms as the number of MDCs/servers increases. We observe that the $N$ algorithm shows high latency across all settings. The $S$ algorithm performs marginally better than $N$, but its performance indicates that scaling alone is not suitable for latency-sensitive systems. On the other hand, the Migration-only policy ($M$) and the combined $SM$ policy are consistently the best-performing algorithms in terms of latency. The $M$ policy improves latency significantly as the number of MDCs increases. Both $M$ and $SM$ show strong improvements when only one server is available, but this improvement does not scale proportionally for two and four servers, which may be due to high coordination overhead.

In Figure~\ref{fig:util_plot_MDCs} and Figure~\ref{fig:util_plot_servers}, we plot the Resource Utilization ($RU$) for all algorithms as the number of MDCs/servers increases. As a general trend, we observe that as resources become more distributed, $RU$ decreases significantly for all algorithms. $SM$ maintains the highest $RU$ across all settings due to its dynamic scaling and migration features. The Scaling-only algorithm $S$ fails to offer a significant improvement compared to the baseline ($N$).

In Figure~\ref{fig:norm_latency_plot_chMult}, we demonstrate the normalised latency for all four algorithms as a function of the charging multiplier ($ch$). The $N$ and $S$ algorithms consistently exhibit the worst latency regardless of the charging rates. For low $ch$ values, $M$ slightly outperforms $SM$; however, as $ch$ increases, $SM$ proves more adaptive to energy changes. This behaviour suggests that $SM$ avoids overloaded resources by using its migration capabilities. Similarly, in Figure~\ref{fig:throughput_plot_chMult}, we analyse the changes in throughput ($TH$) as $ch$ increases. We observe that the Migration-only ($M$) policy maintains a high throughput rate and that the combined $SM$ policy does not significantly increase throughput over $M$ alone. Increasing $ch$ up to 1.5 does not appear promising, as the resulting increase in throughput ($TH$) becomes marginal.

\begin{figure}
\centering
\includegraphics[width=0.45\textwidth]{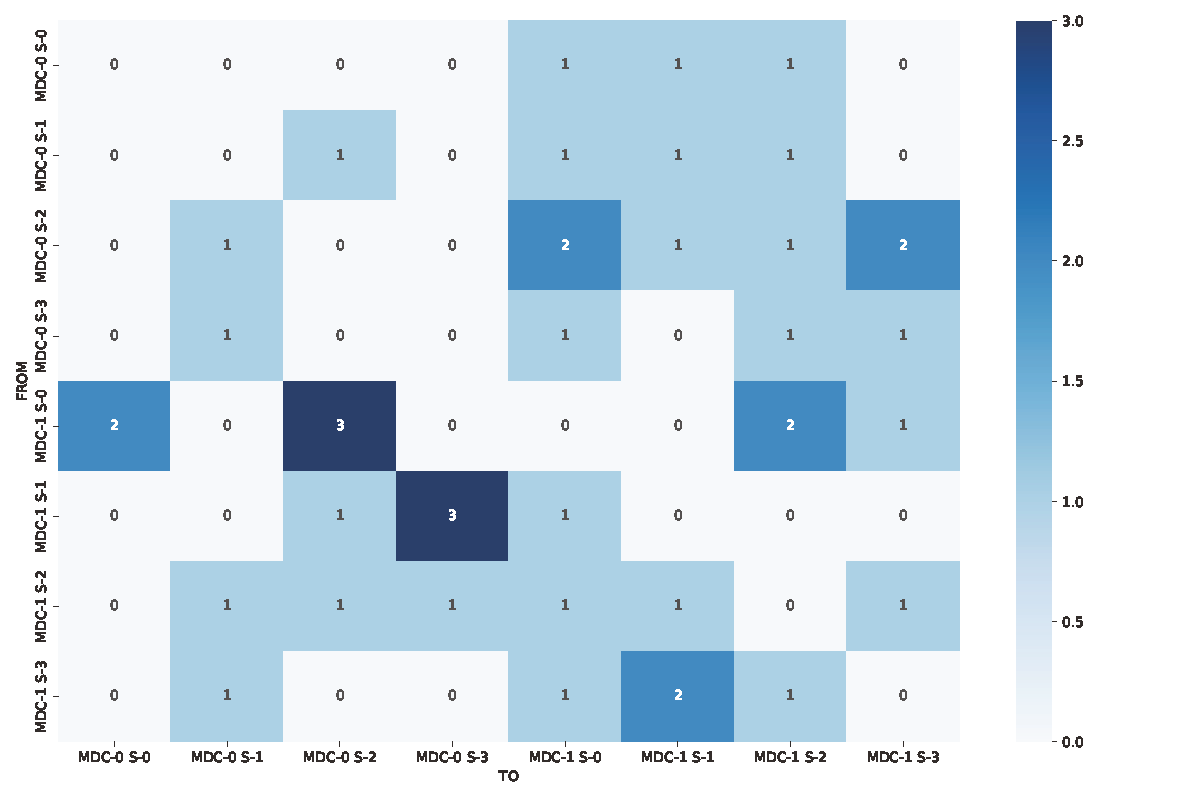}
\caption{Migration Count between Servers}
\label{fig:heatmap_example}
\end{figure}

To conclude the experimental evaluation, this section presents a detailed example of the $SM$ algorithm's operation. For fairness and to better illustrate its behaviour, we intentionally selected a case where $SM$ exhibits moderate performance. In this experiment, two MDCs are used, each equipped with 4 servers and a charging multiplier of 1.25. Figure~\ref{fig:mdc_battery_levels_example} shows the battery levels (i.e., remaining capacity) of both MDCs over time, with horizontal red lines indicating their respective safety thresholds.

Initially, both MDCs are actively utilised by the initial placement algorithm (EDEM). During the first 4000 time steps, latency remains very low (as shown in Figure~\ref{fig:latency_distribution_example}) because both MDCs have sufficient energy to process incoming tasks. However, a critical period arises between time steps 4000 and 6000, during which MDC-2 runs out of energy. According to Figure~\ref{fig:utility}, the energy harvesting capability during this interval is nearly zero, while the task arrival rate increases, causing the energy bottleneck.

As depicted in Figure~\ref{fig:Migrations_example}, most service migrations occur during this low-energy phase, relocating services to MDC-1. This increased load causes MDC-1 to deplete its own battery around time step 8000. Meanwhile, MDC-2 has recovered some energy, prompting the scheduler to migrate services back to it—thereby helping reduce latency once again. Toward the end of the simulation, the demand for processing exceeds the system’s charging capabilities, resulting in a slight increase in overall latency. Finally, the heatmap in Figure~\ref{fig:heatmap_example} demonstrates that the migration policy is effective and well-balanced, as it avoids any "ping-pong" behaviour.

\section{Conclusions and Future Work}\label{sec:con}
This paper introduced a novel online strategy for Multi-access Edge Computing (MEC) systems powered exclusively by renewable energy. Our approach dynamically manages service migration and server frequency scaling to efficiently utilise harvested energy while ensuring low service latency. Experimental results using real-world data demonstrated the effectiveness of our algorithm across various operational scenarios.

We acknowledge several limitations that provide clear directions for future research. Our model assumes accurate forecasting, whereas real-world systems are subject to prediction errors. Besides, our migration model omit overheads such as container image transfer and service cold-start delay, and our battery model is simplified. Our future work will address these limitations directly. We plan to incorporate lightweight, learning-based algorithms to handle forecasting and adapt to dynamic factors like user mobility. Furthermore, we will develop a more comprehensive migration cost model that includes the aforementioned overheads. To ensure a more robust evaluation, we will validate our enhanced strategy using real-world energy harvesting datasets, with the ultimate goal of validating the strategy on a physical testbed.

\section*{Acknowledgements}\label{sec:ack}
This work was supported by the Research Institute of Trustworthy Autonomous Systems (RITAS), Shenzhen Science and Technology Program (No. GJHZ20210
705141807022), Guangdong Province Innovative and Entrepreneurial Team Programme (No. 2017ZT07X386), and SUSTech-University of Leeds Collaborative PhD Programme. Georgios Theodoropoulos is the corresponding author of this article.

\bibliographystyle{unsrt}
\bibliography{ref}

\end{document}